\documentclass[10pt,aps,prl,reprint,twocolumn,amsmath,superscriptaddress,preprintnumbers,floatfix,nofootinbib,showpacs]{revtex4-1}

\usepackage{siunitx}
\usepackage{tikz}
\usepackage{graphicx}
\usepackage{xcolor}
\usepackage[colorlinks=false, pdfborder={0 0 0}]{hyperref}
\usepackage{mathtools}
\usepackage{amssymb}
\usepackage[utf8]{inputenc}
\usepackage[T1]{fontenc}
\usepackage{natbib}

\begin{document}
%\preprint{Draft for PRL \today}
\title{New method for a continuous determination of the spin tune in storage
rings\newline and implications for  precision experiments}
%\author{Editorial team: Dennis Eversmann, Volker Hejny, Fabian Hinder, Jörg  Pretz, Frank Rathmann, Marcel Rosenthal, and Fabian Trinkel}
%\collaboration{JEDI collaboration}
%\affiliation{Institut für Kernphysik, Forschungszentrum Jülich, D-52428 Jülich, Germany}

\author{D.~Eversmann}
\affiliation{III. Physikalisches Institut B, RWTH Aachen University, 52056 Aachen, Germany}
\author{V.~Hejny}
\affiliation{Institut f\"ur Kernphysik, Forschungszentrum J\"ulich, 52425 J\"ulich, Germany}
\author{F.~Hinder}
\affiliation{III. Physikalisches Institut B, RWTH Aachen University, 52056 Aachen, Germany}
\affiliation{Institut f\"ur Kernphysik, Forschungszentrum J\"ulich, 52425 J\"ulich, Germany}
\author{A.~Kacharava}
\affiliation{Institut f\"ur Kernphysik, Forschungszentrum J\"ulich, 52425 J\"ulich, Germany}
\author{J.~Pretz}
\affiliation{III. Physikalisches Institut B, RWTH Aachen University, 52056 Aachen, Germany}
\affiliation{JARA--FAME (Forces and Matter Experiments), Forschungszentrum J\"ulich and RWTH Aachen University, Germany}%
\author{F.~Rathmann}
\affiliation{Institut f\"ur Kernphysik, Forschungszentrum J\"ulich, 52425 J\"ulich, Germany}
\author{M.~Rosenthal}
\affiliation{III. Physikalisches Institut B, RWTH Aachen University, 52056 Aachen, Germany}
\affiliation{Institut f\"ur Kernphysik, Forschungszentrum J\"ulich, 52425 J\"ulich, Germany}
\author{F.~Trinkel}
\affiliation{III. Physikalisches Institut B, RWTH Aachen University, 52056 Aachen, Germany}
\affiliation{Institut f\"ur Kernphysik, Forschungszentrum J\"ulich, 52425 J\"ulich, Germany}
\author{S.~Andrianov}
\affiliation{Faculty of Applied Mathematics and Control Processes, St. Petersburg State University, 198504 Petersburg, Russia}
\author{W.~Augustyniak}
\affiliation{Department of Nuclear Physics, National Centre for Nuclear Research, 00681 Warsaw, Poland}
% ...
\collaboration{JEDI collaboration}

%%%%%%%%%%%%%%%%%%%%%%%%%%%%%%%%%%%%%%%%%%%%%%%%%%%%%%%%%%%%%%%%%%%%%%%%%%%%%%%%%%%%%%%%%%%%%%%%%%%%%%%%%%%%%%%%%%%%%%%%%%%%%%
\begin{abstract}
A new method to determine the spin tune is described and tested. In an ideal planar magnetic ring, the spin tune -- defined as the number of spin precessions per turn -- is given by $\nu_\text{s}=\gamma\cdot G$ ($\gamma$ is the Lorentz factor, $G$ the gyromagnetic anomaly). At $\SI{970}{MeV\per}c$, the deuteron spins coherently  precess at a frequency of $\approx \SI{120}{kHz}$  in the Cooler Synchrotron COSY. The spin tune is deduced  from the up-down asymmetry of deuteron-carbon  scattering. In a  time interval of $\SI{2.6}{s}$, the spin tune was determined with a precision of the order $\num{e-8}$,  and to  $1\times \num{e-10}$ for a continuous $\SI{100}{s}$ accelerator cycle. This renders the presented method a new precision tool for accelerator physics; controlling the spin motion of particles to high precision is mandatory, in particular, for the measurement of electric dipole moments of charged particles in a storage ring.
\end{abstract}

%%%%%%%%%%%%%%%%%%%%%%%%%%%%%%%%%%%%%%%%%%%%%%%%%%%%%%%%%%%%%%%%%%%%%%%%%%%%%%%%%%%%%%%%%%%%%%%%%%%%%%%%%%%%%%%%%%%%%%%%%%%%%%
%%%%%%%%%%%%%%%%%%%%%%%%%%%%%%%%%%%%%%%%%%%%%%%%%%%%%%%%%%%%%%%%%%%%%%%%%%%%%%%%%%%%%%%%%%%%%%%%%%%%%%%%%%%%%%%%%%%%%%%%%%%%%%
\maketitle
%%%%%%%%%%%%%%%%%%%%%%%%%%%%%%%%%%%%%%%%%%%%%%%%%%%%%%%%%%%%%%%%%%%%%%%%%%%%%%%%%%%%%%%%%%%%%%%%%%%%%%%%%%%%%%%%%%%%%%%%%%%%%%
%{\it Introduction}
The matter-antimatter asymmetry that emerges from  the Standard Model (SM) of particle physics falls short by many orders of magnitude compared to the observed value~\cite{doi:10.1146/annurev.nucl.49.1.35}. Physics beyond the SM is thus required and is sought at high energies and by high-precision measurements at lower energies,  for instance in the search for $C\hspace{-.5mm}P$-violating electric dipole moments (EDMs).  %A permanent EDM of a particle or composite non-degenerate system would violate both parity ($P$) and time reversal ($T$), and, via the $C\hspace{-.5mm}PT$ theorem, also $C\hspace{-.5mm}P$ symmetry.
A non-zero EDM measurement would be a telltale sign of physics beyond the SM \cite{Engel201321}. In addition,  EDM measurements of various systems would  point towards the underlying extension  of the SM \cite{Dekens:2014jka,Bsaisou:2014zwa,Bsaisou:2014zwae}.

Up to now, upper limits of hadronic EDMs have been determined for the neutron~\cite{PhysRevLett.97.131801} and the proton, but the latter only indirectly from a measurement on $^\text{199}$Hg~\cite{PhysRevLett.102.101601}. EDMs of charged hadrons are proposed to be  measured in storage rings with a precision of  $10^{-29} \,\text{e} \cdot \text{cm}$ by observing the influence of the EDM on the spin motion~\cite{jedi-collaboration,srEDM-collaboration,Anastassopoulos:2015ura}. The high level of sensitivity is maintained only when the particle spins in the machine precess coherently for long periods of time $(\approx\SI{1000}{s})$. In a series of recent investigations at COSY~\cite{Maier:1997zj,PhysRevSTAB.18.020101}, we  studied  how the spin coherence time of an ensemble of particles can be increased to many hundreds of  seconds through sextupole corrections, bunching, and phase-space cooling of the beam~\cite{guidoboni2013,PhysRevSTAB.17.052803,PhysRevSTAB.15.124202,PhysRevSTAB.16.049901}.

Another limiting factor of the storage ring approach to EDM searches, however,
is  controlling  the spin motion in the presence of small fluctuations of
electric and magnetic fields in order to unambiguously determine the EDM
signal.  Consequently, the measurement described in this paper constitutes one
cornerstone of storage ring EDM searches; {\it viz.}\ the first precise measurement of the spin tune during a complete accelerator cycle.

The spin motion of a particle in the electric and magnetic fields of a  machine  is governed by the Thomas-BMT 
equation~\cite{Thomas:1926dy,Thomas:1927yu,PhysRevLett.2.435},
  extended to include the EDM~\cite{PhysRevLett.2.492,Fukuyama:2013ioa},
\begin{equation}\label{eq:bmt}
     \frac{\text{d}\vec s}{\text{d} t} = \vec s \times \left(\vec \Omega_{\text{MDM}} +  \vec \Omega_{\text{EDM}} \right) \,.
\end{equation}
Here, $\vec s$ denotes the spin vector in the particle rest frame in units of $\hbar$,  
$t$ the time in the laboratory system, and $\vec \Omega_{\text{MDM}}$ and $\vec \Omega_{\text{EDM}}$  
the angular frequencies due to magnetic dipole (MDM) and electric dipole moments (EDM). 
In the following, spin rotations due to EDMs, being many orders of magnitude smaller 
than those produced by MDMs, are neglected. 
It is convenient to define the spin motion relative to the momentum direction~\cite{courrant:1980,lee1997spin}
rotating with the angular velocity $|\vec \Omega_{\mathrm{cyc}}| = qB/(m \gamma)$.
As a result, the spins of particles that orbit in an ideal planar machine  
precess about the vertical magnetic field $\vec B$ relative to the momentum vector with the angular frequency
%\begin{equation}
  $\vec \Omega_{\text{MDM}} = q G \vec B/m$,
  %\label{eq:omega_MDM}\,,
%\end{equation}
where $q$ and $m$ denote particle charge and mass, $G$ is the gyromagnetic anomaly and $\vec B$ the magnetic field at a given point of the particle trajectory. Dividing  $|\vec \Omega_{\text{MDM}}|$ by the cyclotron angular frequency $|\vec \Omega_{\mathrm{cyc}}|$ yields the number of spin revolutions per turn, called the spin tune $\nu_\text{s}$~\cite{lee1997spin,Barber:1998vd}. For a particle on the closed orbit in an ideal magnetic ring,  the spin tune is thus given by
\begin{equation}
    \nu_\text{s} = \gamma G \label{eq:spintune}\,.
\end{equation}
In a real machine, field imperfections, magnet misalignments, and the finite emittance of the beam lead to spin rotations around non-vertical axes and the spin tune deviates from the one given in Eq.~(\ref{eq:spintune}). The most prominent one is the so-called pitch correction~\cite{PhysRevLett.28.1479,Farley197266}.

The experiment was performed at COSY. A  polarized deuteron beam  of  $\approx 10^{9}$  particles  was accumulated, accelerated to the final momentum of $\SI{970}{MeV\per}c$, and electron-cooled to reduce the equilibrium beam emittance. The beam polarization, perpendicular to the ring plane, was alternated from cycle to cycle using two vector-polarized  states, ${p_\xi}^+=0.57\pm 0.01$ and ${p_\xi}^-=-0.49\pm 0.01$, and an unpolarized state. The tensor polarization $p_{\xi \xi}$ of the beam was smaller than 0.02. An rf cavity was used to bunch the beam during the $\approx\SI{140}{s}$ long cycle. After the beam was prepared, the electron cooler was turned off for the remaining measurement period  of $\SI{100}{s}$.

An rf solenoid-induced spin resonance was employed to rotate the spin by $90^\circ$ from the initial vertical direction into the transverse horizontal direction. Subsequently, the beam was slowly extracted onto an internal carbon target using a white noise electric field applied to a stripline unit. Scattered deuterons were detected in scintillation detectors, consisting of rings and bars around the beam pipe~\cite{Albers:2004iw}, and their energy deposit was measured by stopping them in the outer scintillator rings.  The event arrival times with respect to the beginning of each cycle and the frequency of the COSY rf cavity were recorded in one long-range time-to-digital converter (TDC), \textit{i.e.}, the same reference clock was used for all signals. The number of orbit revolutions  could thus be unambiguously assigned to each  recorded event~\cite{PhysRevSTAB.17.052803}.

In the following, we use a right-handed coordinate system, where the $z$-axis points in the beam direction, $y$ upwards, and $x$ sideways. The differential cross section, for scattering of purely vector-polarized deuterons with a vertical polarization component $p_y=0$ off an unpolarized target, can be written as~\cite{PhysRevC.74.064003,PhysRevSTAB.9.050101}
\begin{equation}
      \sigma(\vartheta,\phi) = \sigma_0(\vartheta) \left[1 - \frac{3}{2} p_x(t) A_y^\text{d}(\vartheta) \sin \phi \right]\,.
\label{eq:diff-cs}
\end{equation}
Here, $\sigma_0(\vartheta)$ denotes the differential cross section for unpolarized beam,  $\vartheta$ the polar scattering angle, $\phi$ the azimuthal scattering angle, and $A_y^\text{d}(\vartheta)$ the deuteron vector analyzing power. According to Eq.\,(\ref{eq:bmt}), $p_x(t)$ in Eq.\,(\ref{eq:diff-cs}) oscillates as
\begin{eqnarray}
     p_x(t) =  p_{\xi} \sin( \Omega_\text{s}  t + \varphi)  \label{eq:px(t)}  \,,
\end{eqnarray}
where $\Omega_\text{s}= 2 \pi f_{\mathrm{rev}} \nu_\text{s}$ denotes the angular frequency of the horizontal spin precession,  $\varphi$  the phase, and $p_{\xi}=\sqrt{p_x^2 + p_z^2}$ the magnitude of the in-plane vector polarization. Because of the COSY straight sections, $f_{\mathrm{rev}}$ differs from the cyclotron frequency $f_{\mathrm{cyc}} = \Omega_{\mathrm{cyc}}/(2\pi)$.

In order to determine the spin tune from Eqs.~(\ref{eq:diff-cs}) and (\ref{eq:px(t)}), and to cancel possible acceptance and flux variations during the measurement, asymmetries are formed using the  counts of the Up (U) and Down (D) detector quadrants. The quadrants are centered at $\phi_\text{U}\approx 90^\circ$  and $\phi_\text{D} \approx 270^\circ$, covering polar angles from $\vartheta=9^\circ$ to $13^\circ$,  and an azimuthal range of $\Delta \phi_\text{U} \approx \Delta \phi_\text{D} \approx 90^\circ$. An expression for the  event rate $R_X$ of a  detector quadrant $X=(\text{U}\, \text{or}\,  \text{D})$ is obtained by integration over the solid angle, yielding
\begin{eqnarray}
     R_X &=& I d_\text{t}  \int_X a_X(\vartheta, \phi) \sigma(\vartheta,\phi) \mathrm{d}\Omega \nonumber\\
         &=& I d_\text{t} \overline{\sigma_0}_X  \left( 1- \frac{3}{2} p_x(t) \overline{A^\text{d}_y}_X
%   + \frac{3}{2} p_y \overline{A^{\cos}_y}_X
      \right)  \label{N1}\,.
\end{eqnarray}
Here, $a_X(\vartheta,\phi)$ denotes the combined  detector efficiency and
acceptance, $I$~$[\text{s}^{-1}]$ the beam intensity,
$d_\text{t}$~$[{\text{cm}^{-2}}]$ the target density, $\overline{\sigma_0}_X$
the integrated spin-independent cross section, and
$|\overline{A^\text{d}_y}_X| \approx 0.4 $ the weighted average analyzing power of the respective quadrants.

\begin{figure}[t]
     \includegraphics[width=\columnwidth]{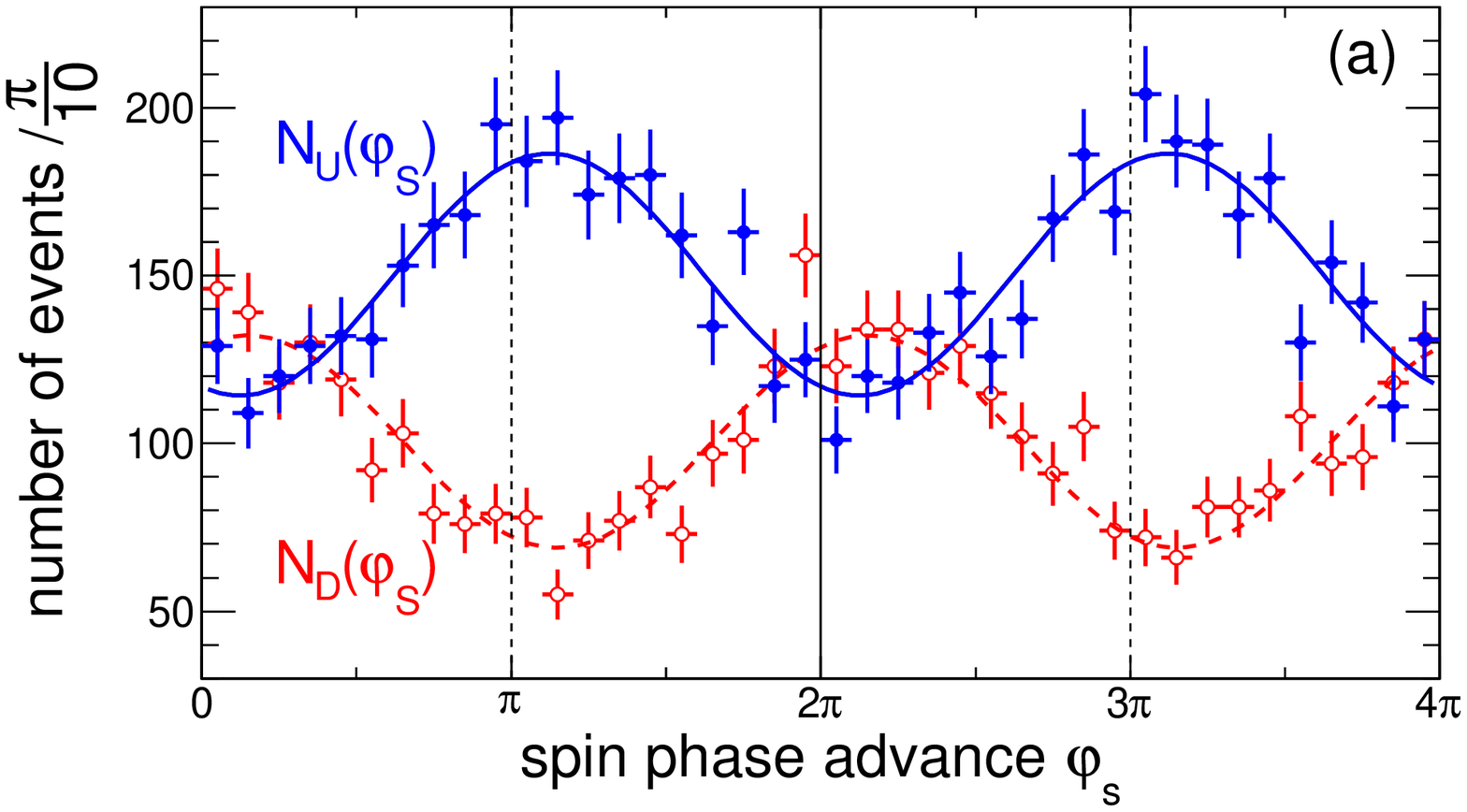}
     \includegraphics[width=\columnwidth]{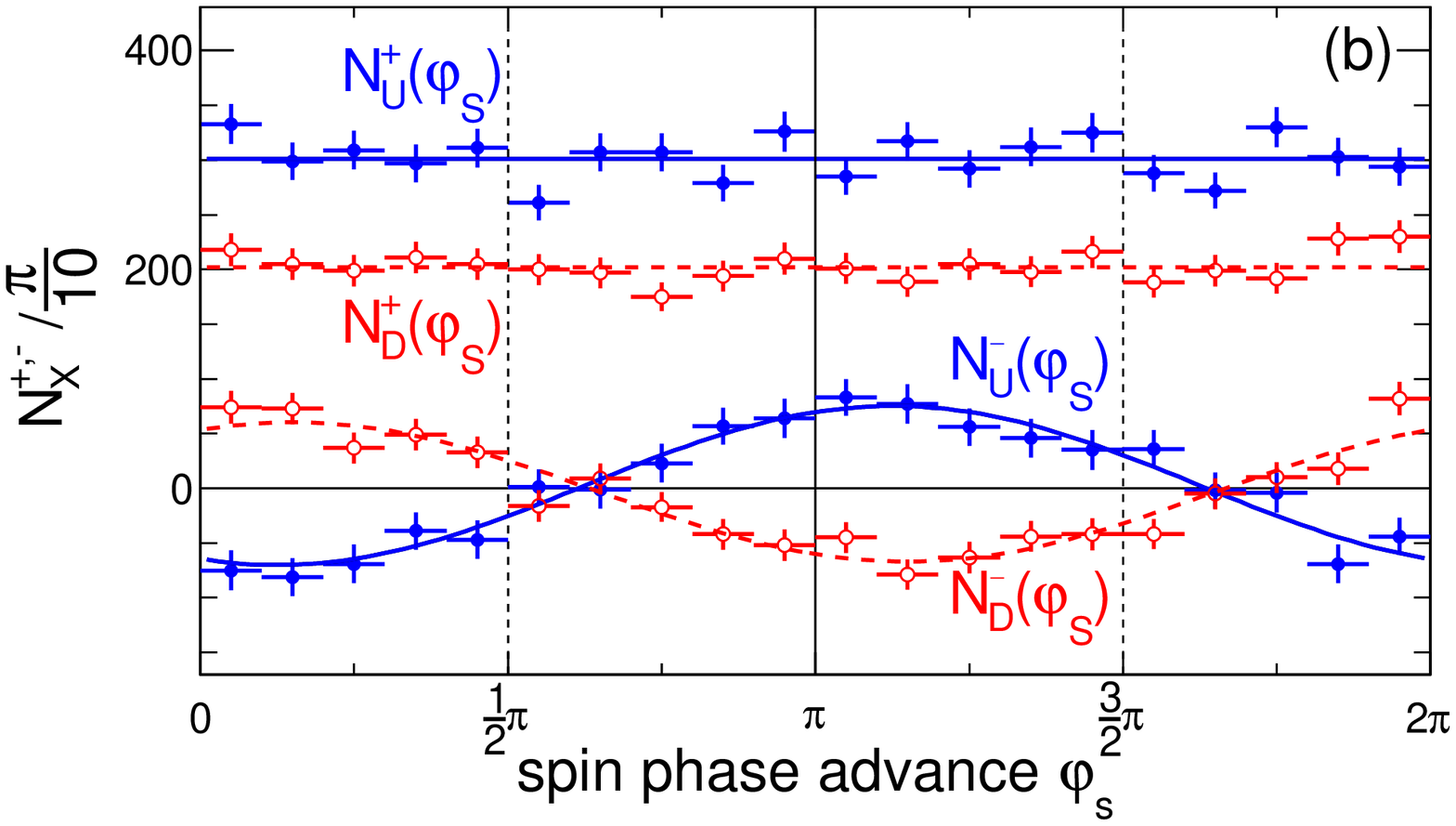}
     \caption{(a): Counts $N_{\text U}$ and $N_{\text D}$ after mapping the events recorded during a turn interval of $\Delta n=10^6$ turns into a spin phase advance interval of $4\pi$. (b): Count sums $N_{\text{U},\text{D}}^+(\varphi_\text{s})$ and differences $N_{\text{U},\text{D}}^-(\varphi_\text{s})$ of Eq.~(\ref{eq:NPlusminus1}) with $\varphi_\text{s} \in [0,2\pi)$ using the counts $N_{\text U}(\varphi_\text{s})$ and $N_{\text D}(\varphi_\text{s})$, shown in panel (a). The vertical error bars show the statistical uncertainties, the horizontal bars indicate the bin width.}
\label{fig:mapping}
\end{figure}

It is not possible  to determine the spin tune $\nu_\text{s}$  from the
observed event rates  by a simple fit with $\nu_\text{s}$ as a parameter using
Eq.~(\ref{N1}), because, at a detector rate of $\approx\SI{5000}{s^{-1}}$ and a spin frequency of $f_\text{s}=|\nu_\text{s}| \cdot f_\text{rev} \approx 0.16 \cdot \SI{750}{kHz} = \SI{120}{kHz}$, 
only about one event is detected per 24 spin revolutions. Therefore, as described below, an algorithm is applied that maps all events into one oscillation period. It generates an asymmetry, largely independent of variations of acceptance, flux, and polarization that oscillates around zero. For each event, the  integer turn number $n$ is calculated, using the event
time compared to the time of the COSY rf cavity. Based on the turn number, the $\SI{100}{s}$ measurement interval  is split into $72$ turn intervals of width  $\Delta n=10^6$ turns (each turn lasting $\approx 1.3 \, \mu \text{s}$). For all events,  the spin phase advance 
$\varphi_\text{s} = 2\pi |\nu_\text{s}^0| n$ is calculated under the assumption of a certain spin tune $\nu_\text{s}^0$. 
In this analysis, we use the absolute value $|\nu_s|$ because the 
asymmetry measured at one location in the ring is insensitive to the sign as
well as to integer offsets of the spin tune.
Each of the turn intervals is analyzed independently, and the events are mapped into a $4\pi$ interval, which yields the event counts $N_{\text U}(\varphi_\text{s})$ and $N_{\text D}(\varphi_\text{s})$  shown in Fig.~\ref{fig:mapping} (a).
In order to obtain from $N_{\text U}(\varphi_\text{s})$ and $N_{\text D}(\varphi_\text{s})$  a sinusoidal wave form that oscillates around zero, four new event counts for the two quadrants ($X=\text{U}\, \text{or}\,  \text{D}$) are defined,
\begin{eqnarray}
 N^{\pm}_\text{X}(\varphi_\text{s}) &=& \label{eq:NPlusminus1}
   \begin{dcases*}
    N_X(\varphi_\text{s}) \pm N_X \left(\varphi_\text{s} + 3\pi\right) & for $0 \le \varphi_\text{s} < \pi$\\
    N_X(\varphi_\text{s}) \pm N_X \left(\varphi_\text{s} + \pi\right)  & for $\pi \le \varphi_\text{s} < 2\pi$\,.
   \end{dcases*}
%   \\ 
% N^{\pm}_\text{D}(\varphi_\text{s}) &=& \label{eq:NPlusminus2}
%   \begin{dcases*}
%    N_\text{D}(\varphi_\text{s}) \pm N_\text{D}\left(\varphi_\text{s} + 3\pi\right) & for $0 \le \varphi_\text{s} < \pi$\\
%    N_\text{D}(\varphi_\text{s}) \pm N_\text{D}\left(\varphi_\text{s} + \pi\right)  & for $\pi \le \varphi_\text{s} < 2\pi$
%   \end{dcases*}
\end{eqnarray}

The above equations provide sums, $N_{\text U}^+(\varphi_\text{s})$ and $N_{\text D}^+(\varphi_\text{s})$, and differences, $N_{\text U}^-(\varphi_\text{s})$ and $N_{\text D}^-(\varphi_\text{s})$, of  counts depicted in Fig.~\ref{fig:mapping} (b).
%\begin{figure}[t]
%     \includegraphics[width=\columnwidth]{pictures/Sorting2.pdf}
%         \caption{Count sums $N_{\text{U},\text{D}}^+(\varphi_\text{s})$ and differences $N_{\text{U},\text{D}}^-(\varphi_\text{s})$ of Eqs.~(\ref{eq:NPlusminus1}, \ref{eq:NPlusminus2}) with $\varphi_\text{s} \in [0,2\pi[$ using the counts $N_{\text U}(\varphi_\text{s})$ and $N_{\text D}(\varphi_\text{s})$, shown in Fig.~\ref{fig:mapping}.}
%\label{fig:sum-diff}
%\end{figure}
While the sums are constant, the differences  oscillate around zero, and the asymmetry,
\begin{eqnarray}
      \epsilon(\varphi_\text{s}) &=& \frac{N_{\mathrm{D}}^{-}(\varphi_\text{s}) - N_{\mathrm{U}}^{-}(\varphi_\text{s})}{N_{\mathrm{D}}^{+}(\varphi_\text{s}) + N_{\mathrm{U}}^{+}(\varphi_\text{s})} \label{eq:eps_updn} \nonumber\\
      &=& \frac{3}{2} p_\xi \frac{\overline{\sigma_{0}}_\text{D} \overline{A^\text{d}_y}_\text{D} - \overline{\sigma_0}_\mathrm{U} \overline{A^\text{d}_y}_\text{U}} {\overline{\sigma_0}_\text{D} + \overline{\sigma_0}_\text{U}} \, \sin(\varphi_\text{s} + \tilde \varphi)\,,
%&=& \tilde \epsilon     \sin(\omega_s \varphi_s+ \varphi) = \tilde \epsilon     \sin(\varphi_s + \varphi) = \epsilon(\varphi_s) \nonumber
\label{eq:eps(phi_s)}
\end{eqnarray}
in the range  $\varphi_\text{s} \in [0,2\pi)$ has the functional form
\begin{equation}
      \label{eq:fitfunc}
      \epsilon(\varphi_\text{s}) = \tilde \epsilon \sin(\varphi_\text{s} + \tilde \varphi)\,,
\end{equation}
independent of beam intensity and target density.
Since the spin coherence time (SCT) of the in-plane vector polarization $p_\xi$ is long ($\tau_\text{SCT}\approx \SI{300}{s}$),
%~\cite{PhysRevSTAB.17.052803}
the polarization is assumed to be constant over  the duration of the turn interval $\Delta n$ ($\SI{1.3}{s}$).

\begin{figure}[t]
     \includegraphics[width=\columnwidth]{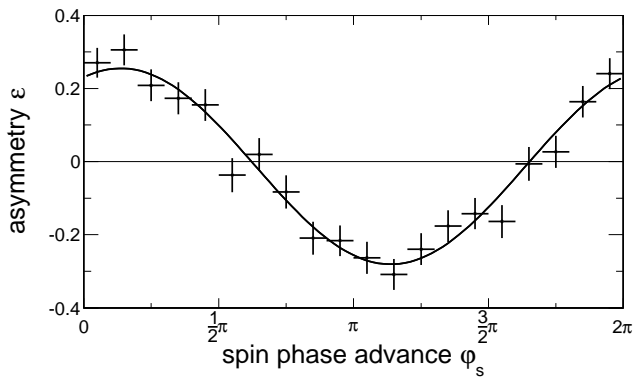}
          \caption{Measured asymmetry $\epsilon(\varphi_\text{s})$ of Eq.~(\ref{eq:eps_updn}) fitted with  $\epsilon(\varphi_\text{s})$ of Eq.~(\ref{eq:fitfunc}) to extract amplitude  $\tilde \epsilon$ and phase $\tilde \varphi$, using  the yields $N_{\text{U}, \text{D}}^{+,-}(\varphi_\text{s})$ of Fig.~\ref{fig:mapping} (b) for a single  turn interval of $\Delta n=10^6$ turns at a measurement time  of $\SI{2.6}{s} < t < \SI{3.9}{s}$.}
\label{fig:fstperiod}
\end{figure}
In every turn interval, the parameters $\tilde \epsilon$ and $\tilde \varphi$ of Eq.~(\ref{eq:fitfunc}) are fitted 
to the measured asymmetry of Eq.~(\ref{eq:eps_updn}). 
An example  is shown in Fig.~\ref{fig:fstperiod}.
The procedure is repeated for several values of $\nu_\text{s}^0$ in 
a certain range around $\nu_\text{s}=\gamma G$ (see \textit{e.g.}, Fig.~5 of~\cite{PhysRevSTAB.17.052803}).

%The fits, for which $\tilde \epsilon$ becomes maximal (an example  is shown in Fig.~\ref{fig:fstperiod}),  yield a first approximation 
%of $\nu_\text{s}$ with a precision of about $10^{-6}$.

%In order to determine the spin tune more accurately, the phase parameter $\tilde \varphi$ is determined 
%from the fits with Eq.~(\ref{eq:fitfunc}) for all turn intervals of a complete cycle. 

\begin{figure}[t]
      \includegraphics[width=\columnwidth]{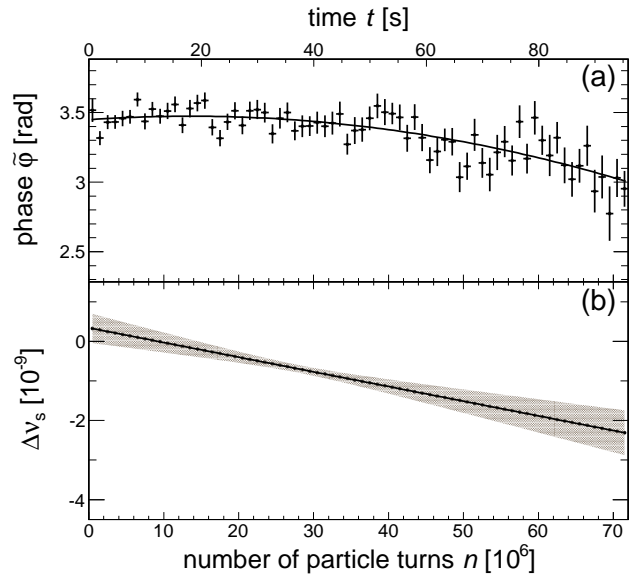}
            \caption{(a): Phase $\tilde \varphi$  as a function of turn number $n$ for all 72 turn intervals of a
            single measurement cycle for $|\nu_\text{s}^\text{fix}| = 0.160975407$, together with a parabolic fit. 
            (b): Deviation $\Delta \nu_\text{s}$ of the spin tune from  $\nu_\text{s}^\text{fix}$ as a function of turn 
            number in the cycle. At $t\approx \SI{38}{s}$, the interpolated spin tune amounts to
            %$\nu_\text{s} = (-16097540771.7 \pm 9.7)\times 10^{-11}$. 
            $|\nu_\text{s}| = (16097540628.3 \pm 9.7)\times 10^{-11}$. 
            The error band shows the statistical error obtained
            from the parabolic fit, shown in panel (a).
            }
\label{fig:phase}
\end{figure}

A fixed common spin tune $|\nu_\text{s}^{\text{fix}}| = 0.160975407$ is chosen such that the phase variation $\tilde \varphi(n)$ 
is minimized, as shown in Fig.~\ref{fig:phase} (a).
The spin tune as a function of  turn number is given by
\begin{equation}\label{eq:nut}
      |\nu_\text{s}(n)| = |\nu_\text{s}^\text{fix}| + \frac{1}{2\pi} \frac{\mathrm{d} \tilde  \varphi(n)}{\mathrm{d}n}
               = |\nu_\text{s}^\text{fix}| + \Delta \nu_\text{s}(n) \,,
\end{equation}
independent of the particular choice of  $\nu_\text{s}^\text{fix}$, because a
different choice for $\nu_\text{s}^\text{fix}$ is compensated for by a corresponding change in $\Delta \nu_\text{s}(n)$.

Without any assumption about the functional form of the phase dependence in Fig.~\ref{fig:phase} (a), one can calculate the spin tune deviation $\Delta \nu_\text{s}(n)$ from $\nu_\text{s}^\text{fix}$ by evaluating ${\mathrm{d}  \tilde  \varphi(n)}/{\mathrm{d} n}$  using two consecutive phase measurements, corresponding to a measurement time of $\SI{2.6}{s}$. At early times  ($\sigma_{\varphi} \approx 0.06$, see Fig.~\ref{fig:phase} (a)) the statistical accuracy of the spin tune reaches $\sigma_{\nu_\text{s}}=1.3 \times 10^{-8}$, and  towards the end of the cycle ($\sigma_{\varphi} \approx 0.15$) $\sigma_{\nu_\text{s}} = 3 \times 10^{-8}$, due to the decreasing event rate.

An even higher precision of the spin tune is obtained by exploiting the observed parabolic phase dependence, fitted to $\tilde{\varphi}(n)$ in Fig.~\ref{fig:phase} (a), which indicates that the actual spin tune changes  linearly as a function of turn number. As displayed in Fig.~\ref{fig:phase} (b), in a single $\SI{100}{s}$ long measurement, the highest precision is reached  at $t\approx \SI{38}{s}$ with an error of the interpolated spin tune of $\sigma_{\nu_\text{s}}=9.7\times 10^{-11}$.

The achieved precision of the spin tune measurements agrees well with  the statistical expectation. The error of a frequency measurement is approximately given by $\sigma_f = \sqrt{6/N} /(\pi \tilde \varepsilon T)$, where $N$ is the total number of recorded events, $\tilde \varepsilon  \approx 0.27$ is the oscillation amplitude of Eq.~(\ref{eq:fitfunc}), and $T$  the measurement duration. In a 2.6 s time interval with an initial detector rate of $\SI{5000}{s^{-1}}$, one would expect an error of the spin tune of  $\sigma_{\nu_\text{s}} = \sigma_{f_\text{s}}/f_\mathrm{rev}\approx  1 \times 10^{-8}$, and, during a $\SI{100}{s}$  measurement with  $N\approx 200000$ recorded events, an error of $\sigma_{\nu_\text{s}} \approx 10^{-10}$.

\begin{figure}[t]
            \includegraphics[width=\columnwidth]{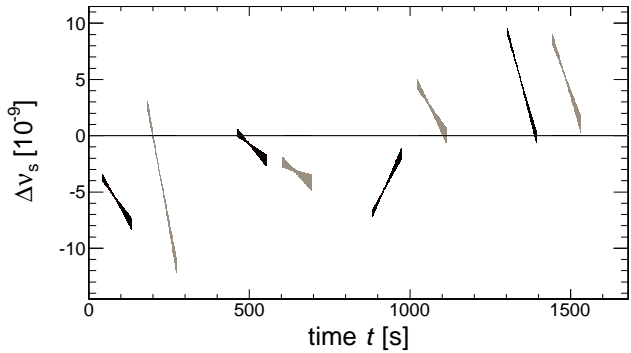}
            \caption{Walk of the spin tune during eight consecutive cycles with alternating  initial vector polarization ${p_\xi}^+$ (black) and ${p_\xi}^-$ (gray). The third cycle is depicted in Fig.~\ref{fig:phase} (b) as well. Cycles with unpolarized beam that followed the  ${p_\xi}^-$ state are not shown.}
            \label{fig:spintunewalk}
\end{figure}

The new method can be used to monitor the stability of the spin tune in the accelerator for long periods of time. As shown in  Fig.~\ref{fig:spintunewalk}, the spin tune variations from cycle to cycle are of the same order ($10^{-8}$ to $10^{-9}$) as those within a cycle [Fig.~\ref{fig:phase} (b)], illustrating that the spin tune determination  provides a new precision tool for the  investigation of systematic effects in a machine. It is remarkable that  COSY is stable to such a precision, because it was not designed to provide  stability below $\approx \num{e-6}$ with respect to, \textit{e.g.}, magnetic fields, closed-orbit corrections  and  power supplies. Presently investigations are  underway to locate the origins of the observed variations in order to develop feedback systems and other means to minimize them further.

Several systematic effects that may affect the spin tune measurement are briefly discussed below. Terms with a vertical vector and a tensor polarization have been omitted in the derivation of $\epsilon(\varphi_\text{s})$ [Eq.~(\ref{eq:eps(phi_s)})]. A detailed analysis taking these terms into account shows that $p_y$ has no influence on the spin tune at all, because the particle ensemble  precesses about the $y$-axis; $p_y$ thus merely dilutes the  asymmetry $\epsilon(\varphi_\text{s})$. Although a small tensor polarization of up to $\pm 0.02$ leads to higher harmonics in the oscillation pattern from which the spin tune is derived (Fig.~\ref{fig:fstperiod}), these contributions alter neither the location of the zero crossings nor that of the extrema, and thus have no influence on the extracted spin tune. In addition, a tilt of the invariant spin axis or misalignment of the detector leads to a modification of the magnitude of the measured asymmetry $\tilde \epsilon$, but neither effect alters the measurement 
of the precession frequency.

Effects of  time-dependent variations of the in-plane polarization, acceptance, and flux  were studied using a Monte Carlo simulation, for which detector rates were generated using Eq.~(\ref{N1}). The  analysis, carried out assuming these quantities to be constant, showed that even extreme variations such as a complete loss of  polarization or acceptance during a $\SI{100}{s}$ measurement, does not affect the spin tune determination   down to a level of $10^{-11}$.

The work presented here can be compared to the measurement of the muon precession frequency $|\vec \Omega_\text{MDM}|$, which was determined in the muon $(g-2)$ experiment with a relative precision of $\approx 10^{-6}$ per year~\cite{PhysRevD.73.072003}. This corresponds to an absolute precision of the spin tune of $\sigma_{\nu_\text{s}} \approx 3 \times 10^{-8}$ per year. The higher precision achieved here is mainly  attributed to the much longer measurement time of $\SI{100}{s}$ compared  to the measurement time of 600\,$\mu$s in the muon $(g-2)$ experiment. Ring imperfections introducing MDM rotations about non-vertical axes make it  impossible at this stage to use the  new technique to directly determine  the gyromagnetic anomaly $G$ with high precision from the measured spin tune.

Future charged particle EDM searches with an anticipated precision of $10^{-29}\,\text{e}\cdot\text{cm}$ can be carried out in frozen-spin mode~\cite{srEDM-collaboration,Anastassopoulos:2015ura}.
%, \textit{i.e.}, with the spins aligned along the momentum of the particles, and clockwise and counter-clockwise beams to cancel out systematic effects.
These investigations, however, demand a new class of storage rings. %providing large electric fields.
Using an existing machine, one could perform a first direct measurement of the proton or deuteron EDM  using an rf Wien filter~\cite{PhysRevLett.96.214802,PhysRevSTAB.16.114001,Rathmann:2013rqa}.  In this case one has to cope with the fast spin precession due to the deflection and  focusing in the magnetic elements. The precision determination of the spin tune allows one to lock the phase of the spin precession to the rf phase of the Wien filter, using a feedback system.

The method to determine the spin tune, described in this paper, provides a precision tool to map out field imperfections, orbit corrections, and beam instabilities. It can be readily extended to protons. In addition, increasing the measurement period by a factor of ten with  $\tau_\text{SCT}$ of a few hundred seconds further increases the  precision of $\nu_\text{s}$ by about the same factor. 

This paper presents  the most precise measurement to date of the spin tune in a storage ring. The current precision reaches a level of $\sigma_{\nu_\text{s}}=\num{e-10}$ for a $\SI{100}{s}$ measurement.
%and in the near future improvements of at least one order in magnitude will become possible.
The new method will have a huge impact on future precision measurements in storage rings, such as
%testing the  $C\hspace{-.5mm}PT$ symmetry and
the determination of electric dipole moments of charged particles.

\begin{acknowledgments}
The authors wish to thank the staff of COSY for providing excellent working conditions and for their support concerning the technical aspects of this experiment. We thank H.O. Meyer for useful discussions. This work has been financially supported by  Forschungszentrum J\"ulich GmbH, Germany, via COSY FFE, the European Union Seventh Framework Programme (FP7/2007-2013) under Grant Agreement No. 283286, and the Shota Rustaveli National Science Foundation of the Republic of Georgia.
\end{acknowledgments}

\bibliographystyle{apsrev4-1}
%\bibliography{newliteratur2}
\bibliography{newliteratur2_short}
%\bibliography{PhysRevLett.2.435} BMT paper
%\bibliography{SpinTune}
%\bibliography{deut_cab}  deuteron carbon scattering
\end{document}